\documentclass[pra,aps,preprint,tightenlines,showpacs,preprintnumbers]{revtex4}
\usepackage{graphics,bm}
\begin{document}

\title{Dynamics of a Bose-Einstein Condensate near a \\ Feshbach Resonance}

\author{R.A. Duine}
\email{duine@phys.uu.nl}
\homepage{http://www.phys.uu.nl/~duine/bec}
\author{H.T.C. Stoof}
\email{stoof@phys.uu.nl}
\homepage{http://www.phys.uu.nl/~stoof}
\affiliation{Institute for Theoretical Physics,
         University of Utrecht, Leuvenlaan 4,
         3584 CE Utrecht, The Netherlands}
\date{\today}

\begin{abstract}
We discuss the response of a Bose-Einstein condensate to a change in the
scattering length, which is experimentally realized by tuning the magnetic
field near a Feshbach resonance. In particular, we consider the collapse
of the condensate induced by a sudden change in the scattering length from
a large positive to a small negative value. We also consider the
condensate dynamics that results from a single pulse in the magnetic
field, due to which the scattering length is rapidly increased from zero
to a large value and then after some time rapidly decreased again to its
initial value. We focus primarily on the consequences of the quantum
evaporation process on the dynamics of the Bose-Einstein condensate, but
also discuss the effects of atom-molecule coherence.
\end{abstract}

\preprint{ITP-UU-02/59}

\maketitle

\def\bx{{\bf x}}
\def\bk{{\bf k}}
\def\half{\frac{1}{2}}
\def\args{(\bx,t)}

\section{Introduction}
\label{introduction}
\begin{figure}
\includegraphics{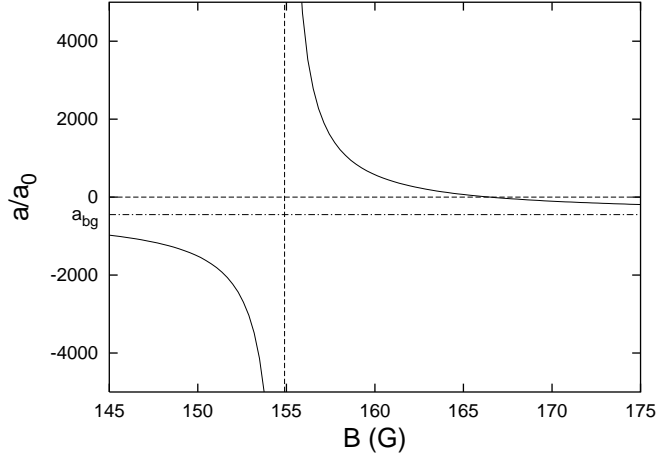}
\caption{\label{resonance}
   The scattering length as a function of the magnetic field for $^{85}$Rb
   in the state $|f=2;m_f=-2 \rangle$. The
   position of the resonance is indicated by the vertical dashed line. At
   the horizontal dashed line the scattering length vanishes. The
   dashed-dotted line indicates the background scattering length
   $a_{\rm{bg}} \approx -450 a_0$.
   }
\end{figure}

From single-channel scattering theory it is well known that the
collisional cross section changes dramatically if the energy of the
incoming particles is close to the energy of a long-lived bound state in
the interaction potential. In particular, the magnitude of the $s$-wave
scattering  length $a$ of an attractive potential well diverges as the
depth of the potential well is increased such that a new bound state
enters the well \cite{sakurai}.  A similar behavior occurs in the case of
a Feshbach resonance, when the energy of the two particles in the incoming
channel is close to the energy of a bound state in a closed channel
\cite{feshbach}. In the case of collisions between alkali atoms, the
coupling between the two channels is provided by the hyperfine
interaction. Due to the spin flips involved in this interaction, the
difference in energy between the bound state and the continuum, the
so-called detuning, is adjustable by means of a magnetic bias field.
Feshbach resonances were first predicted theoretically
\cite{stwalley,eite}, but have now also been observed experimentally, in
various atomic species \cite{inouye,courteille,jake1998,chu}. As a result
the experiments with magnetically-trapped ultracold atomic gases, where
the $s$-wave scattering length fully determines the interaction effects,
have an unprecedented high level of control over the interatomic
interactions. In this paper we focus on $^{85}$Rb in the $f=2,m_f=-2$
hyperfine state, which has a Feshbach resonance at a magnetic field of
$B_0 \approx 154.9$ (G)auss \cite{simon}. Near the resonance the
scattering length, as a function of magnetic field, is given by
\begin{equation}
\label{a_B}
  a (B) = a_{\text{bg} } \left( 1- \frac{\Delta B}{B-B_0} \right).
\end{equation}
The resonance is characterized by the width $\Delta B \approx 11.6$ G and
the off-resonant background scattering length   $a_{\text{bg}} \approx
-450 a_0$, with $a_0$ the Bohr radius. In Fig.~\ref{resonance} the
scattering length is shown as a function of the magnetic field. Clearly,
it can be adjusted experimentally from large negative values, to large
positive ones.  Moreover, at a magnetic field of $B \approx 166.5$ G the
scattering length vanishes, and the gas behaves effectively as an ideal
Bose gas.

With this experimental degree of freedom it is possible to study very
interesting new regimes in the many-body physics of ultracold atomic
gases. The first experimental application was the detailed study of the
collapse of a condensate with attractive interactions. In general a
collapse occurs when the attractive interactions overcome the kinetic
energy of the condensate atoms in the trap. Since the typical interaction
energy is proportional to the density, there is a certain maximum number
of atoms above which the condensate is unstable
\cite{keith1,collapse1,collapse2,marianne,bergeman}. In the first
observations of the condensate collapse by Bradley \textit{et al.}
\cite{curtis1},  a condensate of doubly spin-polarized $^7$Li atoms was
used. These atoms have a magnetic-field independent scattering length of
$a \approx -27 a_0$. For the experimental trap parameters this leads to a
maximum number of condensate atoms that was so low that nondestructive
imaging of the condensate was impossible. Moreover, thermal fluctuations
due to a large thermal component make the initiation of the collapse a
stochastic process, thus preventing also a series of destructive
measurements of a single collapse event \cite{cass1,cass2,rembert2}. A
statistical analysis has nevertheless resulted in important information
about the collapse process. Very recently, it was even  possible to
overcome these problems \cite{jordan}.

In addition to the experiment with $^7$Li, experiments with $^{85}$Rb have
been carried out. In particular, Roberts \textit{et al.} \cite{jake2001}
also studied the stability criterion for the condensate, and Donley
\textit{et al.} \cite{elisabeth} studied the dynamics of a single collapse
event in great detail. Both of these experiments make use of the
above-mentioned Feshbach resonance to achieve a well-defined initial
condition for each destructive measurement. It turns out that during a
collapse a significant fraction of atoms is expelled from the condensate.
Moreover, one observes a burst of hot atoms with an energy of
about $150$ nK. Several mean-field analyses of the collapse, which model
the atom loss phenomenologically by a three-body recombination rate
constant, have offered some theoretical insight
\cite{russen,kerson1999,kerson2000,adhikari,saito,santos2002}. However, the physical
mechanism responsible for the explosion of atoms out of the condensate and
the formation of the noncondensed component is still largely ununderstood
at present.

A second experimental application of the Feshbach resonance has been
implemented  by Claussen \textit{et al.} \cite{neil}. Starting from the
noninteracting limit, the scattering length is made to jump very fast back
and forth to a large and positive value. Surprisingly, in this case one
also observes loss of atoms from the condensate, as well as a burst of hot
atoms. Even more surprising is the fact that the amount of  atoms expelled
from the Bose-Einstein condensate decreases with time during some
intervals, excluding a theoretical explanation in terms of a loss process
characterized by a rate constant.

As a third application Donley \textit{et al.} \cite{elisabeth2} have
conducted an experiment where two trapezoidal pulses in the
magnetic field were applied. As a function of the time between the two pulses an
oscillatory behavior in the number of condensate atoms is observed, which is
attributed to coherent Rabi oscillations between atoms and molecules
\cite{servaas,mackie2002,kohler2002}. 
In this paper we will not consider this experiment, but
instead focus on the first and second experimental applications. In
particular the single-pulse experiment has not received much attention
theoretically, even though an understanding of this experiment seems an
essential first step in the theoretical treatment of the recent two-pulse
experiment. Therefore, the discussion of this most recent experiment will
be postponed to a future publication. 

In a previous paper we have considered the loss of atoms by means
of elastic two-body collisions, in the situation where the condensate
collapses \cite{rembert1}. However, the mechanism put forward in this
paper  is much more general. In particular, the mechanism should also be
relevant for the above mentioned single-pulse experiments. The main goal
of this paper is to present the theory behind it in great detail. Since
the mechanism is able to describe loss from a condensate at zero
temperature we will hereafter refer to it as quantum evaporation. The
two-pulse experiments of Donley {\it et al.} \cite{elisabeth2} have made it
clear that atom-molecule coherence can have an important effect on the
dynamics of a Bose-Einstein condensate. Besides the quantum evaporation
process, we thus want to consider this physics in the case of the
single-pulse experiments as well. We are able to achieve this because  
very recently we have derived an effective quantum field theory that
offers a description of the Feshbach resonance in terms of an
atom-molecule hamiltonian that captures all the relevant two-body physics
exactly \cite{rembert4}. Apart from a detailed discussion of the
condensate collapse, the main application of this paper therefore concerns
the effect of quantum evaporation in the single-pulse experiments and the
investigation of the importance of atom-molecule coherence in this case. With
respect to the latter remark it should be noted that the effect of
atom-molecule coherence in the case of the condensate collapse will be
neglected in the following, because the magnetic field is tuned to a far
off-resonant value to induce the collapse. 

In view of this the above the paper is organized as follows. In
Sec.~\ref{quantum_evaporation} we present and discuss the theoretical
description of quantum evaporation. In Sec.~\ref{applications} we present
the applications to the condensate collapse, and to the single-pulse
experiments with positive scattering length. We end in
Sec.~\ref{conclusions} with our conclusions.

\section{Quantum Evaporation}
\label{quantum_evaporation}
In this section we derive the generalized Gross-Pitaevskii equation that
includes the correction term due to the quantum evaporation of atoms out
of the condensate. From this result follows a rate equation for the change
in the number of atoms in the condensate. Sec.~\ref{subsec:ggpe} is rather
technical and may be omitted in a first reading. To facilitate this the
final rate equation, which is most important for our purposes, is
presented in Sec.~\ref{subsec:rateeqn}.

\subsection{Generalized Gross-Pitaevskii equation}
\label{subsec:ggpe}
Although the desired rate equation for the number of atoms can also be derived
from the imaginary-time formalism by means of a Wick rotation to real
time, the equation of motion for the condensate wave function can not be
derived in this manner.  Therefore, we use a functional formulation of the
Schwinger-Keldysh nonequilibrium theory \cite{schwinger,keldysh} developed
in Refs.~\cite{henk1,henk2,henkltp}, from which the equation of motion for
the condensate wave function follows directly as the equation for the
``classical'' part of the fluctuating order-parameter field.

Within this formalism, the Wigner probability distribution of the order
parameter is written as a functional integral over complex fields $\psi^*
\args$ and $\psi \args$. These fields are defined on the Keldysh contour
${\mathcal C}^t$, which runs from $t_0$ to $t$ and then back to $t_0$.
The probability distribution is given by
\begin{equation}
\label{probdistr}
  P[\phi^*,\phi;t] = \int_{\psi^* \args = \phi^* (\bx)}^{\psi \args = \phi
  (\bx) }
   d [\psi^*] d [\psi] \exp
  \left\{\frac{i}{\hbar} S[\psi^*,\psi]
  \right\},
\end{equation}
where we absorbed the appropriate initial condition $P[\phi^*,\phi;t_0]$
into the measure of the functional integral \cite{henkltp}.
The action in the exponent of the integrand is given by
\begin{eqnarray}
\label{action}
  S[\psi^*,\psi] = \int_{{\mathcal C}^t} d t' \int d \bx'
  \psi^*(\bx',t')
    \left[
   i \hbar \frac{\partial}{\partial t'}+\frac{\hbar^2 \nabla^2}{2m}
    - V^{\text{ext}} (\bx')  - \frac{T^{\text{2B}} (t')}{2}
   |\psi (\bx',t')|^2   \right] \psi (\bx',t'),
\end{eqnarray}
where $V^{\text{ext}}(\bx)$ is the external trapping potential. The
interaction is determined by the two-body T(ransition) matrix element
$T^{\text{2B}} (t)=4 \pi a (t) \hbar^2/m$, where $a(t) \equiv a(B(t))$ is the
interatomic $s$-wave scattering length, and $m$ is the mass of one atom.
Note that we explicitly allowed the scattering length to depend on time.
This is experimentally realized by tuning the magnetic field near the
Feshbach resonance.

To arrive at an effective action for the condensate wave function, we
explicitly separate out the condensate part from the field
$\psi(\bx,t)$. Therefore we write
$\psi(\bx,t)=\psi_0(\bx,t)+\psi'(\bx,t)$, and substitute this into the
action in Eq.~(\ref{action}). In this separation, $\psi_0 (\bx,t)$
describes the condensed part of the gas, whereas $\psi'(\bx,t)$ describes
the fluctuations. The precise distinction between the condensate and the
noncondensed part is
discussed in detail in Sec~\ref{applications}. Physically, the idea is that
$\psi_0 \args$ describes the Bose-Einstein condensate and its collective
modes, whereas $\psi' \args$ is associated with the modes not occupied by the
condensate. To define these two parts
consistently, we have to require that they are essentially orthogonal, i.e.,
\begin{equation}
\label{orthogonalcond}
  \int d \bx \left[  \psi_0^*(\bx,t) \psi' (\bx,t) + \psi'^* (\bx,t)
  \psi_0
  (\bx,t) \right]= 0.
\end{equation}
This condition ensures that the Jacobian of the transformation of integration
variables in the functional integral in Eq.~(\ref{probdistr}) is equal to
one. In the operator formalism, this condition
implies that the Bose field operators $\hat \psi'(\bx,t)$ and $\hat
\psi'^{\dagger} (\bx,t)$ associated
with the fluctuations, obey the usual Bose commutation relations in the
Fock space built upon the states orthogonal to $\psi_0 \args$.

After this substitution the functional integral becomes
\begin{eqnarray}
\label{probdistrnew}
   P[\phi^*, \phi;t] &=&\int d [\psi_0^*] d [\psi_0]
   \exp \left\{
   \frac{i}{\hbar} S[\psi_0^*,\psi_0] \right\} \nonumber \\
  &\times& \int d[\psi'^*] d[\psi'] \exp \left\{
  \frac{i}{\hbar}   S_{\text{int}} [\psi_0^*,\psi_0,\psi'^*,\psi']
  + \frac{i}{\hbar} S'[\psi_0^*,\psi_0,\psi'^*,\psi']
  \right\}.
\end{eqnarray}
Here, we define $S'[\psi_0^*,\psi_0,\psi'^*,\psi']$ such that
it contains the terms up to quadratic order in the fluctuations.  We do
not retain the terms proportional to $(\psi')^2$ and $(\psi'^*)^2$, since
these so-called anomalous terms are only needed to describe the collective motion of the
condensate. Their effect is  therefore already included in the action $S[\psi_0^*,\psi_0]$
for the part of the field that
describes the condensate. In principle $S_{\text{int}}
[\psi_0^*,\psi_0,\psi'^*,\psi']$ contains terms which are either of first,
third or fourth order in the fluctuations. We neglect, however, the terms
of third and fourth order, which is known as the Bogoliubov approximation.
This approximation is valid for the zero-temperature applications under
consideration in this paper. Moreover, the terms of higher order in
the fluctuations describe the interactions among the ejected atoms and are
expected to be of little importance in determining the ejection rate
for the atoms expelled from the condensate.

We write the quadratic action $S'[\psi_0^*,\psi_0,\psi'^*,\psi']$ as
\begin{eqnarray}
\label{sbogo}
  S'[\psi_0^*,\psi_0,\psi'^*,\psi'] &=&
    \int_{{\mathcal C}^t} d t'
    \int d \bx'
    \int_{{\mathcal C}^t} d t''
    \int d \bx''
     \psi'^* (\bx',t')
     \hbar G^{-1} (\bx',t';\bx'',t'')
     \psi' (\bx'',t''),
\end{eqnarray}
where we introduced the Green's function $G(\bx,t,\bx',t')$ for
the fluctuations by means of
\begin{eqnarray}
\label{greensfunction}
   \left[ i \hbar \frac{\partial}{\partial t}
    + \frac{\hbar^2 \nabla^2}{2m} - V^{\text{ext}} (\bx)
    -2 T^{\text{2B}} (t) |\psi_0(\bx,t)|^2 \right] G(\bx,t,\bx',t')
      = \hbar \delta (\bx-\bx') \delta (t,t').
\end{eqnarray}
Here, the delta function in the time variables is defined on the Keldysh
contour, by means of $\int_{{\mathcal C}^t} dt' \delta (t,t') =1$.
The part of the action that describes the interactions between the condensed
and noncondensed parts of the system is, in first instance, given by
\begin{eqnarray}
\label{sintold}
   && S_{\text{int}} [\psi_0^*,\psi_0,\psi'^*,\psi']
    = \int_{{\mathcal C}^t} d t' \int d \bx'
  \nonumber \\
    &&\qquad \qquad \left\{ \psi_0^* ( \bx', t')
      \left[ i \hbar \frac{\partial}{\partial t'}
      +\frac{\hbar^2 \nabla^2}{2m} - V^{\text{ext}} (\bx')
        - T^{\text{2B}} (t') |\psi_0 (\bx',t')|^2
      \right] \psi' (\bx',t') \right. \nonumber \\
     &&\qquad \qquad+\left. \psi'^* ( \bx', t')
    \left[  i \hbar \frac{\partial}{\partial t'}
            +\frac{\hbar^2 \nabla^2}{2m} - V^{\text{ext}} (\bx')
          - T^{\text{2B}} (t') |\psi_0 (\bx',t')|^2
              \right] \psi_0 (\bx',t') \right\}.
\end{eqnarray}
It is important to note that this part of the action does not vanish
because the condensate wave function, as we see in a moment, does not obey
the usual Gross-Pitaevskii equation once we include the quantum
evaporation process. Furthermore, the field $\psi' \args$ describes
the high-energy part of the system, and thus has an expansion in terms of
the high-energy trap states. As a result we are allowed to neglect the terms
proportional to the single-particle Hamiltonian. The terms with the time
derivative vanish because of the orthogonality condition in
Eq.~(\ref{orthogonalcond}). The action in Eq.~(\ref{sintold})
therefore reduces to
\begin{equation}
\label{sint}
  S_{\text{int}} [\psi_0^*,\psi_0,\psi'^*,\psi']
   = -\int_{{\mathcal C}^t} d t' \int d \bx' \ \left[
   J^* (\bx',t') \psi' (\bx',t') +
   \psi'^* (\bx',t') J (\bx',t') \right],
\end{equation}
where we introduced the ``current density''
\begin{eqnarray}
\label{defj}
 J \args=
     T^{\text{2B}} (t) |\psi_0 \args|^2 \psi_0 \args.
\end{eqnarray}

The functional integral over the fluctuations in Eq.~(\ref{probdistrnew})
is a Gaussian integral and can be easily performed. This defines
the effective action for the condensate on the Keldysh contour by means of
\begin{eqnarray}
   P[\phi^*, \phi;t] &=&\int d [\psi_0^*] d [\psi_0]
   \exp \left\{
   \frac{i}{\hbar} S[\psi_0^*,\psi_0] \right\} \nonumber \\
  &\times& \int d[\psi'^*] d[\psi'] \exp \left\{
  \frac{i}{\hbar}   S_{\text{int}} [\psi_0^*,\psi_0,\psi'^*,\psi']
  + \frac{i}{\hbar} S'[\psi_0^*,\psi_0,\psi'^*,\psi']
  \right\} \nonumber \\
  &\equiv& \int d [\psi_0^*] d [\psi_0]
   \exp \left\{ \frac{i}{\hbar} S_{\text{eff}} [\psi_0^*,\psi_0] \right\},
\end{eqnarray}
and results in
\begin{eqnarray}
\label{seff}
  S_{\text{eff}} [\psi_0^*,\psi_0]
  &=& S[\psi_0^*,\psi_0] \nonumber \\ && -\frac{1}{\hbar}
   \int_{{\mathcal C}^t} dt' \int d \bx' \int_{{\mathcal C}^t} dt''
   \int d \bx'' \
    J^* (\bx',t') G (\bx',t';\bx'',t'')
      J (\bx'',t'').
\end{eqnarray}
Because the Green's function in this effective action is equal to
\begin{equation}
\label{defgreens}
  i G (\bx',t';\bx'',t'') \equiv \text{Tr} \left\{ \hat
  \rho (t_0) T_{{\mathcal C}^t}
 \left(
     \hat \psi' (\bx',t')
     \hat \psi'^{\dagger} (\bx'',t'')
  \right) \right\}_{J=0},
\end{equation}
where $T_{{\mathcal C}^t}$ denotes time ordering along the Keldysh contour
and $\hat \rho (t_0)$ represents the initial density matrix of the gas,
this Green's function can be decomposed into its analytic pieces by means
of
\begin{equation}
\label{decompositiongf}
  G (\bx,t;\bx',t') = \theta (t,t') G^{>} (\bx,t;\bx',t')
                     + \theta (t',t) G^{<} (\bx,t;\bx',t'),
\end{equation}
with $\theta (t,t')$ the Heaviside function on the Keldysh contour. The
Green's functions $G^{>}$ and $G^{<}$ thus correspond to averages of a
fixed order of creation and annihilation operators. Note that these
Green's functions essentially describe the propagation of a noncondensed
atom in the presence of the mean-field interaction with the condensate.

To finally derive the generalized Gross-Pitaevskii equation for the
condensate wave function, we want to separate out the ``classical'' part
of the field $\psi_0 \args$. This is achieved by means of the
transformation
\begin{equation}
\label{separation}
  \psi_0 (\bx, t_{\pm}) = \phi \args \pm \frac{\xi \args}{2}.
\end{equation}
Here, $\phi \args$ denotes the condensate wave function, whereas the field
$\xi \args$ describes its fluctuations. The upper (lower) sign in
Eq.~(\ref{separation}) corresponds to the forward (backward) branch of the
Keldysh contour. When substituting this transformation into the effective
action in Eq.~({\ref{seff}}), we should in principle only keep terms up to
quadratic order in the fluctuations, to avoid a double counting of the
interactions that we have already taken into account. However, to read off
the generalized Gross-Pitaevskii equation, including the correction terms
associated with the quantum evaporation process, it suffices to consider
only the linear terms in the fluctuations. This is can be understood from
the fact that with this approximation a functional integration over the
fluctuations leads to a constraint for $\phi \args$, which is precisely
the ``classical'' equation of motion that we are interested in. With the
transformation in Eq.~(\ref{separation}) we thus project the effective
action on the real-time axis and read off the equations of motion for
$\phi \args$ and $\phi^* \args$ by putting the coefficient of the terms
linear in $\xi^* \args$ and $\xi \args$ equal to zero, respectively. After
straightforward but somewhat tedious algebra this results in
\begin{eqnarray}
\label{gpeqnelastic}
 i \hbar \frac{\partial \phi \args}{\partial t} &=&
    \left[
      -\frac{\hbar^2 \nabla^2}{2m}
      + V^{\text{ext}} (\bx)  + T^{\text{2B}} (t)
      |\phi (\bx,t)|^2
    \right] \phi \args \nonumber \\ &&
    + \left\{ \frac{T^{\text{2B}}(t)}{\hbar}
         \int_{-\infty}^{\infty} dt' \int d \bx' T^{\text{2B}} (t')
        \left[ 2 \phi^* \args G^{(+)} (\bx,t;\bx',t') \phi (\bx',t')
         \right. \right. \nonumber \\
          && \left. \left. +\phi \args  G^{(-)} (\bx',t';\bx,t) \phi^* (\bx',t')
    \right] |\phi (\bx', t')|^2 \rule{0mm}{6mm} \right\} \phi \args
\end{eqnarray}
with the complex conjugate expression for $\phi^* (\bx,t)$.
We defined the retarded and the advanced Green's
functions in the usual way by
\begin{equation}
\label{defgreensretadv}
  G^{(\pm)} (\bx,t;\bx',t') = \pm \theta (\pm (t-t'))
            \left[ G^{>} (\bx,t;\bx',t') - G^{<} (\bx,t;\bx',t')
        \right].
\end{equation}
Note that the Heaviside function in this definition
is precisely such that the equation of motion for $\phi \args$ is causal.

\begin{figure}
\includegraphics{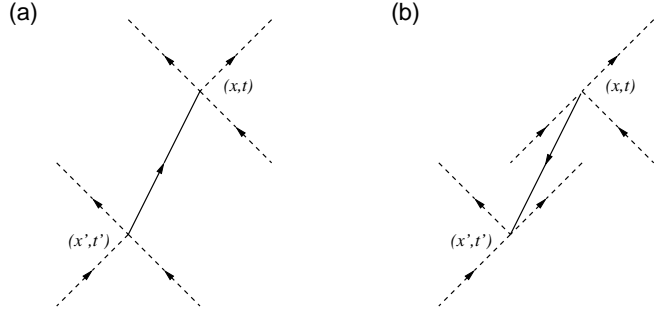}
\caption{\label{feynman}
         These time-ordered Feynmans diagrams correspond to
     terms associated with quantum evaporation
      in the equation of motion for the condensate wave
        function
        }
\end{figure}
The time-ordered Feynman diagrams corresponding to the two terms in the
generalized Gross-Pitaevskii equation describing the quantum evaporation
process are given in Figs.~\ref{feynman}~(a)~and~(b), respectively. In these
diagrams, a condensate atom is denoted by a dashed line. An ingoing dashed
line thus corresponds to a factor $\phi \args$, and an outgoing dashed
line to a factor $\phi^* \args$. The advanced and retarded propagators of
the ejected atoms are denoted by solid lines. A retarded propagator is
denoted by an arrow pointing from $(\bx',t')$ to $\args$, and corresponds
to the propagation of a particle. An advanced propagator is denoted by an
reversed arrow, and can be interpreted as the propagation of a ``hole''.
Note that the first term in Eq.~(\ref{gpeqnelastic}) has an additional
factor of two with respect to the second term. This is understood from the
fact that the corresponding diagram (Fig.~\ref{feynman}~(a)) has two
outgoing lines at $\args$ and therefore contributes twice, whereas
the diagram in Fig.~\ref{feynman}~(b) has only one outgoing line at those
coordinates.

\subsection{Rate equation}
\label{subsec:rateeqn}
In the previous section we have generalized the Gross-Pitaevskii equation to
include the quantum evaporation process. With this equation we derive a
rate equation for the number of atoms in the condensate. It is given by
\begin{eqnarray}
\label{rateeqn}
  \frac{d N_{\text{c}} (t)}{dt} &=& \frac{d}{dt} \int d \bx |\phi \args|^2 =
   \frac{2}{\hbar^2} \int d \bx \int_{-\infty}^{\infty} dt' \int d \bx'
   \nonumber \\
   &&\times {\text{Im}}
   \left[
   T^{\text{2B}} (t) |\phi \args|^2 \phi^* \args
   G^{(+)} (\bx,t;\bx',t')
   T^{\text{2B}} (t') |\phi (\bx', t')|^2 \phi
  (\bx',t')
  \right].
\end{eqnarray}
The retarded propagator $G^{(+)} (\bx,t;\bx',t')$ describes the
noncondensed atoms.  Note that in both this rate equation, and the
equation of motion in Eq.~(\ref{gpeqnelastic}) we have taken the limit
$t_0 \to -\infty$ to eliminate initial transient effects. Furthermore, we
have made use of the fact that $\left[G^{(+)}
(\bx,t;\bx',t')\right]^*=G^{(-)} (\bx',t';\bx,t)$, which follows from the
definition of this Green's function.

The nonmarkovian rate equation in Eq.~(\ref{rateeqn}) describes the
evolution of the number of condensate atoms due to a change in the
scattering length or the condensate wave function. Physically the
coupling between the condensed and noncondensed parts of the system allows
the condensate to eject atoms. Since we have neglected the
interactions among the ejected atoms, this rate equation is applicable
only for short times and does not describe the rethermalization of the
ejected atoms. Because of this approximation the ejection process is coherent,
and atoms can come back into the condensate on short time scales. To get
more physical insight in the quantum evaporation process, we discuss the
weak-coupling limit, where it just corresponds to elastic condensate
collisions.

In the weak-coupling limit, where the mean-field interaction is small
compared to the energy-level splitting of the external trapping
potential,  the rate equation for the number of atoms in the condensate
acquires a familiar form. To see this, we note that in this
limit the retarded propagator is given by
\begin{equation}
  G^{(+)} (\bx,t;\bx',t') = -i \theta (t-t') \sum_{{\bf n} \neq 0}
  e^{-\frac{i}{\hbar} \epsilon_{{\bf n}} (t-t')} \chi_{{\bf n}} (\bx)
  \chi^*_{{\bf n}} (\bx'),
\end{equation}
because the single particle states $\chi_{{\bf n}} (\bx)$ and energies
$\epsilon_{\bf n}$ are, in this limit,
by definition not affected much by the interactions. The sum in this
expression is over all the trap states, except for the condensate mode,
i.e., the one-particle ground state.

The condensate wave function is given by
\begin{equation}
  \phi \args = e^{-\frac{i}{\hbar} \mu t} \sqrt{N_{\text{c}} (t)} \chi_0 (\bx),
\end{equation}
where $\mu \equiv \epsilon_0$ is the ground state energy, and $N_{\text{c}}
\gg 1$ is the number of
atoms in the condensate.
Using these expressions, the rate
equation for the number of atoms in Eq.~(\ref{rateeqn}) can be rewritten as
\begin{equation}
\label{rateeqnfgr}
   \frac{d N_{\text{c}} (t)}{dt}
     = -\left[ \frac{N_{\text{c}}^3 (t)}{2} \right]  \frac{2 \pi}{\hbar}
     \sum_{{\bf n} \neq 0}
                    \delta (\epsilon_{\bf n} - \epsilon_{0})
                       \left| \langle {\bf n} |\hat V| 0 \rangle
               \right|^2,
\end{equation}
which is precisely Fermi's Golden Rule for the rate to scatter out of the
initial state $| 0 \rangle$, found from second-order perturbation
theory. The sum is over all final states of the form
\begin{equation}
  \langle \bx_1 \bx_2 | {\bf n} \rangle \equiv \frac{1}{\sqrt{2}}
    \left[ \chi_0 (\bx_1) \chi_{\bf n} (\bx_2) +
           \chi_{\bf n} (\bx_1) \chi_0 (\bx_2)
    \right],
\end{equation}
with energy $\epsilon_{\bf n}+\mu$. Since we are dealing with identical
bosons these states are symmetric. This final state thus represents a
condensate atom and an ejected atom, whereas the initial state with energy
$2 \mu$ is given by
\begin{equation}
  \langle \bx_1 \bx_2 | 0 \rangle \equiv \chi_0 (\bx_1) \chi_{0}
  (\bx_2),
\end{equation}
and therefore represents two condensate atoms. The additional factor
$N_{\text{c}}^3$ is a result of the Bose statistics of the atoms. There is
a factor $N_{\text{c}}(N_{\text{c}}-1)/2 \approx N_{\text{c}}^2/2$ for
the number of condensate atom pairs and an additional factor $1+N_{\text{c}}
\approx N_{\text{c}}$ for the condensate atom that is Bose stimulated back
into the condensate. Finally, the potential is given by
\begin{equation}
  \hat V=T^{\text{2B}} \delta (\hat \bx_1-\hat \bx_2).
\end{equation}

\begin{figure}
\includegraphics{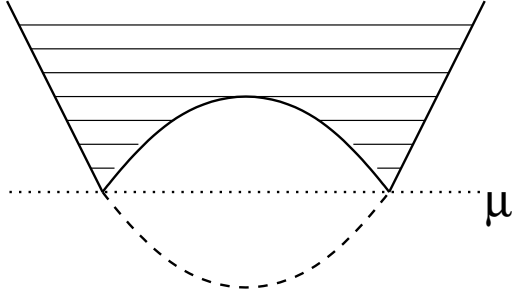}
\caption{ \label{energy_conservation}
   Energy levels of the noncondensate atoms (thin solid lines), in the
   presence of a condensate at a given chemical potential $\mu$ (dotted
   line). The potential felt by the noncondensed atoms is indicated by the
   solid line.
  }
\end{figure}

We see that, in the weak-coupling limit, the energy-conserving delta
function can never be obeyed, and thus $dN_{\text{c}}/dt=0$, as expected
in this limit. Another important observation is that energy conservation
also forbids the ejection of atoms out of a static condensate with
repulsive interactions, even in the strong-coupling limit. This is because
of the fact that the energy of a condensate atom, i.e., the chemical
potential, is positive in this case, and that the energy of the excited
states is always larger than the chemical potential. In the Thomas-Fermi
limit, where the mean-field interaction is much larger than the kinetic
energy of the atoms, this is illustrated in
Fig.~\ref{energy_conservation}. In this figure, the harmonic trapping
potential is denoted by the dashed line. For a given chemical potential
$\mu$, indicated by the dotted line, the noncondensed atoms feel the
potential indicated by the solid line. The energy levels of the states in
this potential are also drawn schematically. Clearly, the energy of the
first excited state is always larger than the chemical potential, and
therefore the ejection of atoms out of the condensate by means of elastic
collisions is forbidden due to energy conservation.

This leads to the important conclusion that in order to eject atoms out of
the condensate by means of the quantum evaporation process, there has to
be a strong time dependence of either the condensate wave function, or of
the scattering length. In the next section, we discuss two experimentally
relevant examples where this is the case.

\section{Applications}
\label{applications}
In this section we apply the theory, described in the previous section, to
two different experimental situations. In the first part, we
consider the case where the Feshbach resonance is used to investigate the
collapse of a condensate \cite{simon,elisabeth}. In the second part we
apply our theory to the situation where the scattering length is changed
very rapidly back and forth from zero to a large and positive value \cite{neil}.
We also discuss the importance of atom-molecule coherence in this case.

\subsection{Condensate collapse}
The use of a Feshbach resonance has made it possible to explore the
physics of the collapse in two different regimes. First, we can have a
collapse in which the dynamics of the condensate is mostly determined by
the mean-field interactions. This is the case in the experiments with
$^7$Li \cite{curtis1,cass1,cass2,jordan}, which has a fixed negative
scattering length. In these experiments, there is
always a large thermal component which feeds the condensate. Therefore, we
 call such a collapse a Type I collapse, in analogy with a Type I
supernova, which is believed to be the result of an accreting white dwarf
that explodes when the accumulated mass becomes to large \footnote{More
precisely, such a supernova is called a Type Ia supernova. The proper
classification scheme of supernovae is based on the characteristics of
their spectra.}, similar to the $^7$Li condensate that grows from the
thermal cloud and then collapses. The more recent experiments by Donley
\textit{et al.} \cite{elisabeth} also deal with a Type I collapse, since
one starts a collapse from the noninteracting limit, where the density is
relatively high compared to Bose-Einstein condensates with a repulsive
self-interaction.

Complementary to this regime we can have a second type of collapse where
the dynamics is mostly determined by the external trapping potential. This
regime was considered experimentally by Cornish \textit{et al.}
\cite{simon}. Using the Feshbach resonance of $^{85}$Rb one  first makes a
large, stable and essentially pure condensate, i.e., there is no visible
thermal cloud present, with repulsive interactions. Then, one suddenly
switches the interactions from repulsive to slightly attractive, and watches the
subsequent collapse. Because the collapse in this case starts at low
density due to the initially large and repulsive interatomic interactions
and the magnitude of the final negative scattering length is much smaller
than the initial positive one, its dynamics is mostly determined by the
trapping potential. We call such a collapse a Type II collapse, in
analogy with a Type II supernova. Such a supernova is the fate of a
massive star analogous to the large condensate with repulsive
interactions.

In the experiments on the Type II collapse it was found that after such a
collapse the number of atoms in the condensate is of the same order as the
maximum number of atoms allowed to have a metastable condensate
\cite{simonpc}. This suggests that during such a collapse enough atoms are
ejected from the condensate, so that the condensate becomes metastable.
The most important theoretical task is therefore to identify the physical
mechanism responsible for this ejection. We have recently argued that the
quantum evaporation process provides an explanation for these experiments
\cite{rembert1}. Moreover, we will explicitly show here that the decay
solely by means of three-body recombination does not explain the
experimental results for the Type II collapse.

Interestingly, the metastable condensate that is the result of a Type II
collapse is precisely the starting point for a Type I collapse, making
this last collapse a much more violent phenomenon. We restrict ourselves
here to the description of the Type II collapse, because a proper
treatment of the Type I collapse requires a full numerical solution
of the generalized Gross-Pitaevskii equation that includes the nonlocal
correction term due to quantum evaporation. This is beyond the scope of
the present paper, and we use a simpler variational approach here that we
believe is appropriate for a Type II-collapse event.

\subsubsection{Gaussian approximation and semiclassical retarded propagator}
In principle, we must now numerically solve the generalized Gross-Pitaevskii equation
in Eq.~(\ref{gpeqnelastic}) that includes nonlocal terms.
However, in order to gain physical insight, we will make several
approximations to reduce the solution of this equation to a numerically
more tractable problem. First of all, we use a gaussian variational
approach to the condensate wave function
\footnote{In Ref.~\protect\cite{rembert1} we used the form $\phi \args =
     1/(2 \pi)^3 \int d \bk \ e^{-\frac{i}{\hbar} \epsilon (\bk) t} \phi
     (\bk,t)$, where $\phi (\bk,t)$ is the Fourier transform of the
     gaussian \textit{ansatz}, and assumed the energy of a condensate
     atoms to be negligibly small with respect to the energy of an
     ejected atom.  Moreover, in the retarded propagator of the
     ejected atoms we neglected the mean-field energy with respect to
     the kinetic energy of the ejected atoms. Here we go beyond these
     approximations.
     }.
More precisely, we assume the condensate wave function to be of the form
\begin{equation}
\label{gaussianansatz}
 \phi ({\bf x},t) = \sqrt{N_{\text{c}} (t)} e^{i \theta_0 (t)}
  \prod_j \left( \frac{1}{\pi q_j^2(t)} \right)^{1/4}
  \times \exp \left\{ - \frac{x_j^2}{2q_j^2(t)}
                      \left( 1 - i \frac{m q_j(t)}{\hbar}
                                       \frac{d q_j(t)}{dt} \right)
         \right\}~.
\end{equation}
In the limit of a small number of condensate atoms $N_{\text{c}} (t)$ this
\textit{ansatz} becomes an exact solution of the Gross-Pitaevskii equation
for a harmonic external trapping potential, and therefore a good
description of the condensate after the collapse. It is, however, also
known that a gaussian \textit{ansatz} gives good results on the frequencies
of the collective modes, even in the Thomas-Fermi regime \cite{perez}.
Moreover, since we consider only the Type II collapse and therefore by
definition assume that the trapping potential is more important than the
mean-field interactions, we expect the gaussian \textit{ansatz} to give
physically reasonable results at all times.

In the case of the ordinary Gross-Pitaevskii equation the variational
parameters obey Newton's equation of motion \cite{collapse2}
\begin{equation}
\label{eom}
m \frac{d^2 q_j(t)}{dt^2}
       = - \frac{\partial}{\partial q_j} V({\bf q}(t);N_{\text{c}}(t))~,
\end{equation}
with a potential given by
\begin{equation}
\label{pot_q}
V({\bf q};N_{\text{c}}) =
 \sum_j \left( \frac{\hbar^2}{2m q_j^2} + \frac{m \omega_j^2 q_j^2}{2} \right)
 + \sqrt{\frac{2}{\pi}} \frac{a (t) \hbar^2 N_{\text{c}}}{m q_x q_y q_z},
\end{equation}
where the frequencies $\omega_j$ are the frequencies of the harmonic
external trapping potential in the three spatial directions. We have
recently extended the above variational calculus also to the case of a
nonlinear Schr\"odinger equation including an imaginary term due to the
presence of a thermal cloud,  and found that the equation of motion for
the variational parameters in principle contains a damping term
\cite{rembert2}. However, we are interested in the zero-temperature
situation, in which this damping term is negligible.  This means that for
a description of the collapse, we only have to couple Eq.~(\ref{eom}) to
the rate equation for the number of atoms in Eq.~(\ref{rateeqn}).

To determine this rate equation in detail, we need an expression for the
propagator of the ejected atoms. Anticipating that the energy of the
ejected atoms will be much higher than the energy of a condensate atom, we
take for the one-particle Green's function of the ejected atoms the
semiclassical approximation,
\begin{equation}
\label{greensfluc}
  G^{(+)} (\bx,t;\bx',t') = -i \theta (t-t') \int \frac{d {\bf k}}{(2 \pi)^3}
     e^{-\frac{i}{\hbar} \epsilon ({\bf k},{\bf R},T) (t-t')}
     e^{i {\bf k} \cdot (\bx-\bx')},
\end{equation}
where we introduced the center of mass coordinate ${\bf R}=(\bx+\bx')/2$,
and used $T=(t+t')/2$ for notational convenience. The energy of the
ejected atoms is, in this approximation, given by
\begin{equation}
\label{energyejected}
  \epsilon ({\bf k},{\bf R},T) = \frac{\hbar^2 {\bf k}^2}{2 m} +
  V^{\text{ext}}
  ({\bf R}) + 2 T^{\text{2B}} (T) |\phi ({\bf R},T)|^2 + \epsilon_1.
\end{equation}
This energy is measured from $\epsilon_1$, which is the first excited
level in the trap. This energy thus determines the cut-off between the
condensate and the ejected atoms. Although this cut-off is important to
make the distinction between condensate and noncondensate atoms, we find
that our numerical results are not very sensitive to its precise value.

Since the most important contribution to the rate equation comes from
the center of the trap, we expect that we can, to a good approximation,
take for the energy of an ejected atom the value $\epsilon ({\bf k},{\bf
0},T) \equiv \epsilon ({\bf k},T)$. We use this value for the energy
of an ejected atom from now on. This approximation is very convenient,
since we can now perform the spatial integrals in the rate equation, as
well as the integral over the momenta in the expression for the retarded
propagator analytically. The final result is given in
Appendix~\ref{app:rateeqn}.

In the \textit{ansatz} in Eq.~(\ref{gaussianansatz}) we have included a
global phase $\theta_0 (t)$. This is important, since we have already seen
in the previous section that this global phase determines the energy of a
condensate atom in equilibrium. We can determine this global phase by
insertion of the \textit{ansatz} in the Gross-Pitaevskii equation including
the correction term in Eq.~(\ref{gpeqnelastic}), and separating out the
real part. Neglecting the contribution of the quantum evaporation process
to the mean-field energy, it is up to an irrelevant constant given by
\cite{rembert2}
\begin{eqnarray}
  \theta_0 (t) = -\frac{1}{\hbar} \left\{
  \int_{-\infty}^t d \tau \left[ \frac{\partial V ({\bf q}
  (\tau),
  N_{\text{c}} (\tau))}{\partial N_{\text{c}}}
  - \sum_j \frac{1}{4} m \dot q_j^2 (\tau)
  \right] + \sum_j \frac{1}{4} m q_j (t) \dot q_j (t) \right\}.
\end{eqnarray}
Note that in equilibrium this results in a phase factor $e^{-i \mu
t/\hbar}$ for the condensate wave function, with $\mu=\partial V({\bf
q},N_{\text{c}})/\partial N_{\text{c}}$. This is clearly the correct
expression for the chemical potential.

With the above approximations we have reduced the difficult problem of
solving a partial differential equation with a nonlocal correction term to
the problem of solving the ordinary differential equations for the
variational parameters in Eq.~(\ref{eom}) coupled to the rate equation for
the change in the number of atoms in Eq.~(\ref{rateeqn}). This rate
equation is now only nonlocal in time, since all the integrals can be done
analytically for the gaussian wave function and the semiclassical retarded
propagator we are considering here. In the next section we will present
our results obtained by numerically solving these equations.

\subsubsection{Results}
We have performed numerical simulations for the experimental conditions of
Cornish \textit{et al.} \cite{simon,simonpc} with $^{85}$Rb. The frequencies
of the external trapping potential are equal to $\omega_r/2 \pi=17.4$ Hz
and $\omega_z/2 \pi=6.8$ Hz, in the radial and axial direction,
respectively. One starts with a condensate consisting of $N_{\text{c}} (0)
\approx 4000$ atoms and a scattering length of $a(B_{\text{i}})=2500 a_0$,
where $B_{\text{i}}$ is the magnetic field at the initial time. Initially, there is
no visible thermal cloud present so the system is approximately at zero
temperature. One then ramps the magnetic field linearly in a time $\Delta
t$ from its initial value $B_{\text{i}}$ to the final value $B_{\text{f}}$,
chosen such that $a(B_{\text{f}})=-60 a_0$. The ramp time is taken equal to
$\Delta t=0.5$ ms.

\begin{figure}
\includegraphics{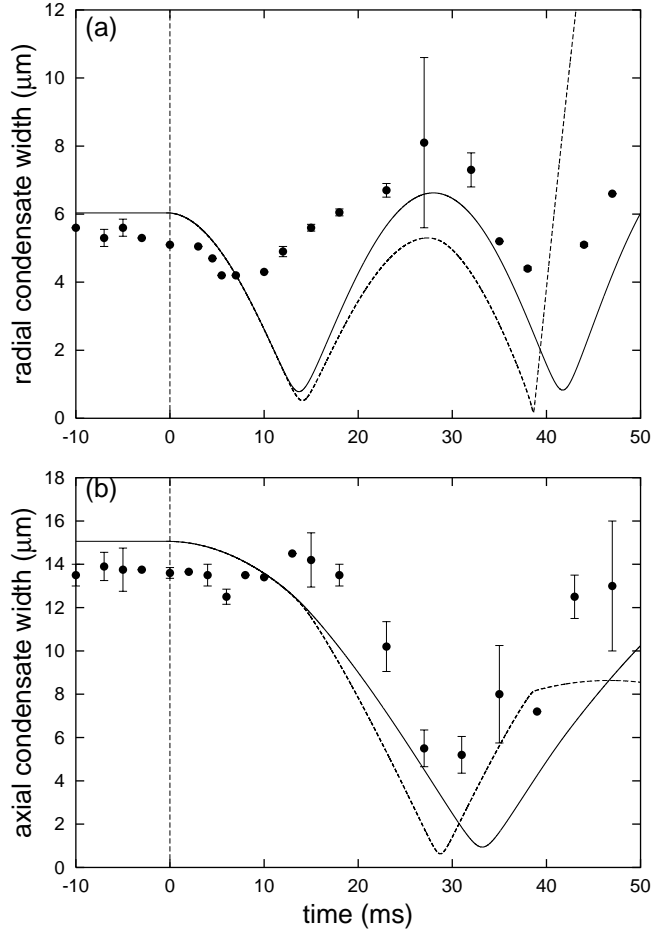}
\caption{ \label{q_t}
  (a) Radial and (b) axial width of the condensate as a function of time
  during a Type II collapse. At the origin of the time axis the scattering
  length vanishes as it is changed from large and postive to negative. The
  solid line corresponds to a simulation where both quantum evaporation
  and three-body recombination are included. The dashed line corresponds
  to a simulation including only three-body recombination. The
  experimental points are also shown and taken from Ref.~\protect\cite{simonpc}.
  }
\end{figure}

To investigate the importance of three-body recombination during the
collapse, we have included it in our simulations in addition to the quantum
evaporation process. This amounts to including a term
\[
   -\frac{i \hbar}{2} \frac{K_3}{3!} |\phi \args|^4 \phi \args
\]
on the right hand side of Eq.~(\ref{gpeqnelastic}), from which the
contribution to the rate equation for the number of atoms can be easily
found. Although there are several predictions for the normal-component
rate constant $K_3$ \cite{moerdijk,fedichev,esry,braaten}, its behavior
as a function of the magnetic field is unknown near the Feshbach resonance
and precise experimental data is unavailable \cite{jake2000}. Following
Saito and Ueda \cite{saito}, we take $K_3=1.2 \times 10^{-27}$ cm$^6$/s.
With this value these authors have been able to explain some of the
results of the experiment on the Type I collapse by Donley \textit{et al.}
\cite{elisabeth}. Since the final value of the magnetic field is in the
same range for both experiments, we expect the three-body recombination
rate constant to be of the same order of magnitude.

\begin{figure}
\includegraphics{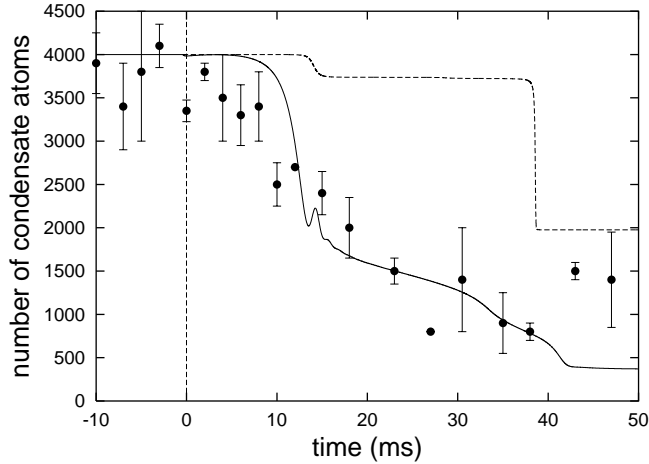}
\caption{\label{n_t}
   Number of atoms in the condensate as a function of time during a
   Type II collapse. The solid line corresponds
  to a simulation where both quantum evaporation and three-body
  recombination are included. The dashed line corresponds to a simulation
  including only three-body recombination. The experimental points are
  also shown and taken from Ref.~\protect\cite{simonpc}.
   }
\end{figure}

In Figs.~\ref{q_t}~and~\ref{n_t} we present the results of our
simulations. The widths of the condensate in the radial and axial
direction, i.e., the variational parameters $q_r$ and $q_z$, are shown as
a function of time in Fig.~\ref{q_t}~(a) and (b), respectively. The number
of atoms as a function of time is displayed in Fig.~\ref{n_t}. The origin
of the time axis is chosen such that the scattering length is equal to
zero at $t=0$. The solid line corresponds to a simulation that includes
both the quantum evaporation process and three-body recombination. The
condensate collapses first in the radial direction in a time $\pi/(2
\omega_r) \approx 14$ ms and during the last part of this radial collapse
the condensate ejects a large fraction of its atoms, by means of the
quantum evaporation process. In our simulations, we find that the
three-body recombination hardly contributes. After a time $\pi/(2
\omega_z) \approx 36$ ms the axial width of the condensate reaches its
minimum, resulting again in a slight increase of the ejection rate. Note
that these time scales are expected, since the dynamics is determined by
the external trapping potential. More remarkably, for a very short time
the number of atoms increases with time during the collapse. In principle,
this can occur, because the quantum evaporation process is coherent, and
can thus describe multimode Rabi oscillations between the condensate and
the noncondensed part of the system, as we shall see in much more detail
in the next section.

The simulation that includes the quantum evaporation process shows
quantitative agreement with the experimental results \cite{simonpc}, which
are also shown in the Figs.~\ref{q_t} and \ref{n_t}. The disagreement for
small values of the widths of the condensate is a result of  the fact that
the experimental resolution for the condensate size is about $4$ $\mu$m
\cite{simonpc}.

We have also performed a simulation that includes only three-body
recombination and no quantum evaporation. The results for this simulation
are shown in Figs.~\ref{q_t} and \ref{n_t} by the dashed lines. The fact
that the minima in the widths of the condensate in Fig.~\ref{q_t} are
lower than the results including quantum evaporation, indicates that the
condensate density has to be relatively high for recombination of atoms to
occur. Once such a high density is reached, the ejection of atoms occurs
very fast, resulting in a staircase-like pattern for the number of atoms
as a function of time, which is clearly not visible in the experimental
data. At $t \approx 38$ ms the condensate decays so fast, that the radial
direction becomes stable again, resulting in a large increase of the
radial width.

Finally we make some remarks about the properties of the ejected atoms. In
Ref.~\cite{rembert1} we have calculated the distribution of the kinetic
energy of the ejected atoms, as well as the angular distribution. We have
also performed preliminary calculations of the average kinetic
energy emitted in the radial and axial directions, and have found that the
energy distribution of the ejected atoms is anisotropic. Donley \textit{et
al.} \cite{elisabeth} have measured the angular and radial temperatures of
the emitted atoms and have indeed found an anisotropic distribution of the
energies. However, our calculations are done for the Type II collapse and
our approximations are especially suited for this case, whereas these
experiments deal with a Type I collapse. Therefore, we do not directly
compare our results with the available experimental data of Donley
\textit{et al.} \cite{elisabeth}. Moreover, to improve the experimental
resolution limit, one expands the gas at the end of each destructive
measurement by an increase in the scattering length. Therefore, the
resulting mean-field energy and rethermalization effects may play an
important role in determining the energy of the ejected atoms, which is
not included in our calculations of the energy distribution function in
Ref.~\cite{rembert1}.

In conclusion, we have shown in this subsection that the simulations of
the Type II collapse that include quantum evaporation show quantitative
agreement with the experimental results. The most important feature of
these experimental results is the fact that the condensate starts to eject
atoms almost immediately after the initiation of the collapse. We have
also shown that solely three-body recombination does not account for this
rapid onset of the loss of atoms. At this point it is important to notice
that this conclusion also holds if we numerically solve the generalized
Gross-Pitaevskii equation, without approximations. The reason for this is
that the gaussian \textit{ansatz} used here is certainly appropriate for
the first part of the collapse, when the dynamics is not yet very violent.
This is borne out by numerical simulations of the Gross-Pitaevskii
equation, which  have also shown that when the highest densities are
reached during a Type I collapse, high-density ``spikes'' can form on the
profile of the wave function
\cite{kerson1999,kerson2000,adhikari,saito,santos2002}.
These ``spikes'' are not included in our \textit{ansatz} for the
condensate wave function and therefore our approximations might be less
appropriate for the highest densities reached during the collapse.
However, our results suggest that such high-density ``spikes'' may well
never occur if one includes the effects of the quantum evaporation
process.

\subsection{Multimode Rabi oscillations}
\begin{figure}
\includegraphics{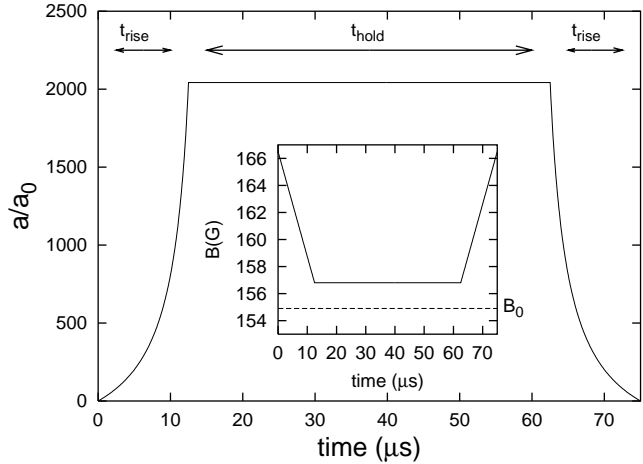}
\caption{ \label{pulse}
     The scattering length as a function of time in the experiments by
     Claussen \textit{et al.} \protect\cite{neil}, and the corresponding
     magnetic field (inset). One increases the scattering length in a time
     $t_{\text{rise}}$ by means of a linear ramp in the magnetic field, and
     holds the magnetic field for a time $t_{\text{hold}}$, before ramping
     back. The position of the Feshbach resonance is indicated by the
     dashed line.
  }
\end{figure}

Apart from the negative scattering length regime, the experimental control
over the interatomic interactions has also made it possible to explore the
regime where the interaction is large and positive. To this end, Claussen
\textit{et al.} \cite{neil} have conducted an experiment where the
magnetic field undergoes a trapezoidal pulse in time, resulting in a quick
jump in the scattering length towards a large positive value. In detail,
one ramps the magnetic field linearly in a time $t_{\text{rise}}$ from its
value in the noninteracting limit ($B \approx 166.5$ G), to  a  value
where  the scattering is of the order of a few thousand Bohr radii.  The
magnetic field is kept at this value for a time $t_{\text{hold}}$ before
ramping back to the initial value again within the same $t_{\text{rise}}$.
The scattering length as a function of time for a typical pulse is shown
in Fig.~\ref{pulse} and the inset shows the corresponding magnetic field.
The rise time $t_{\text{rise}}$ and the hold time $t_{\text{hold}}$ are
typically of the order of $10-100$~$\mu$s.

In this experiment, one observes particle loss from the condensate as a
function of the both the rise time and the hold time, accompanied by a
``burst'' of atoms from the condensate. The temperature of the burst atoms
is of the same order as in the case of the experiments on the collapse,
i.e., about $150$ nK. Most importantly, the amount of atoms lost from the condensate
decreases with increasing $t_{\text{rise}}$ over some interval, which can never be
the case for atom loss characterized by a rate constant, such as
three-body recombination. Also note that the time scales of the pulse are
very small compared to the time scales set by the inverse frequencies of
the external trapping potential. This means that the density profile of
the condensate hardly changes its shape during the pulse, since the
collective modes which alter the density profile have frequencies on the
same order of magnitude.

\subsubsection{Retarded propagator}
The single-pulse experiments start in the noninteracting limit where the
density profile of the condensate has the shape of a gaussian, and the
pulse in the magnetic field is so fast that the condensate wave function
hardly changes due to the interactions. This makes it convenient to expand
the propagator for the ejected atoms in the excited harmonic oscillator
eigenstates, since the condensate part is then left out most easily. The
fact that the condensate density profile almost remains the shape of the
ground state of the trap makes this problem easier to deal with
theoretically and makes it therefore worthwhile to determine the
propagator for the ejected atoms as accurate as possible. With the
Type~II-collapse problem of the previous subsection this objective is much
more difficult since then the condensate wave function changes its shape
considerably during the collapse. This means that at each time a different
number of trap states has to be included in the wave function, making the
cut-off between condensate and noncondensed atoms strongly time-dependent.
Therefore we have applied several approximations in that case. Even though
the density profile of the condensate does not change much, the phase of
the condensate wave function does change very fast in the single-pulse case.
Therefore we use for the phase of condensate wave function the phase of
the gaussian \textit{ansatz} given in Eq.~(\ref{gaussianansatz}). Since
this wave function also describes the low-lying collective modes of the
condensate, we have to exclude these from the propagator of the
noncondensed atoms.

Denoting the eigenstates of the single-particle hamiltonian again by $ \chi_{
\bf n} (\bx)$ and the corresponding eigenvalues by $\epsilon_{\bf n }$, we
have for the propagator of the ejected atoms
\begin{equation}
\label{greensfctpulse}
  G^{(+)} (\bx,t;\bx',t') = -i \theta (t-t') {\sum_{\bf n,n'}}'
   a_{\bf n,n'} (t,t') \chi_{ \bf n} (\bx) \chi^*_{ \bf n'} (\bx')
    e^{-\frac{i}{\hbar} (\epsilon_{\bf n} t - \epsilon_{\bf n'}
    t')}.
\end{equation}
The prime denotes summation over all the excited trap states not contained
in the condensate wave function. The
equation of motion for the expansion coefficients $a_{\bf n,n'} (t,t')$
can be found by inserting Eq.~(\ref{greensfctpulse}) into the equation of
motion for the Green's function given in Eq.~(\ref{greensfunction}).
However, it is easier to realize that the Green's function is in
Eq.~(\ref{defgreens}) shown to be related to the expectation value of the
product of two Heisenberg annihilation and creation operators for the
noncondensed atoms. The equation of motion for the annihilation operator
of interest is given by
\begin{eqnarray}
\label{heisenbergeom}
  \left[ i \hbar \frac{\partial}{\partial t}  + \frac{\hbar^2
  \nabla^2}{2m} - V^{\text{ext}} (\bx) - 2 T^{\text{2B}} (t)|\phi \args|^2
  \right] \hat \psi' \args = 0,
\end{eqnarray}
with the hermitian conjugate expression for the creation operator. We
solve this equation by expanding the annihilation operator as
\begin{equation}
\label{expansionoperator}
  \hat \psi' \args = {\sum_{\bf m}}' \phi_{\bf m} \args \hat \psi'_{\bf m},
\end{equation}
where the Schr\"odinger operator $\hat \psi'_{\bf m}$ annihilates an atom
in the harmonic-oscillator state with quantum number ${\bf m}$. We then
expand also the functions $\phi_{\bf m} \args$ in trap states by means of
\begin{equation}
\label{expansionfunction}
  \phi_{\bf m} \args = {\sum_{\bf n}}' c_{\bf n}^{\bf m} (t)
  e^{-\frac{i}{\hbar} \epsilon_{\bf n} t}\chi_{\bf n} (\bx),
\end{equation}
and determine the equation of motion for the coefficients $c_{\bf n}^{\bf m} (t)$
from Eq.~(\ref{heisenbergeom}). This results in
\begin{equation}
\label{eomcoeffs}
  \frac{d c_{\bf n}^{\bf m} (t)}{dt} =
   -\frac{2i}{\hbar} T^{\text{2B}} (t) {\sum_{\bf n'}}' V_{\bf n,n'} (t)
   c_{\bf n'}^{\bf m} (t)  e^{-\frac{i}{\hbar} (\epsilon_{\bf
   n'}-\epsilon_{\bf n}) t},
\end{equation}
with matrix elements given by
\begin{equation}
\label{matelnts}
  V_{\bf n,n'} (t) = \int d \bx \chi^*_{\bf n} (\bx) |\phi
  \args|^2
  \chi_{\bf n'} (\bx),
\end{equation}
which depend on time through the variational parameters in the gaussian
\textit{ansatz} in Eq.~(\ref{gaussianansatz}) and the number of condensate
atoms. These matrix elements can be calculated analytically and the result is
given in Appendix.~\ref{app:matrixelnts}. The advantage of the above
approach is that we do not have to solve the equation for the derivative
of $a_{\bf n,n'} (t,t')$ with respect to $t'$ separately.

Putting the results together, we find for the coefficients in the
expansion of the Green's function the expression
\begin{equation}
\label{expanscoeffs}
  a_{\bf n,n'} (t,t') = {\sum_{\bf m}}' c^{\bf m}_{\bf n} (t)
   c^{\bf m}_{\bf n'} (t').
\end{equation}
With this Green's function we have performed simulations of the single-pulse
experiments by Claussen \textit{et al.} \cite{neil}, of which the results are
presented in the next section.

\begin{figure}
\includegraphics{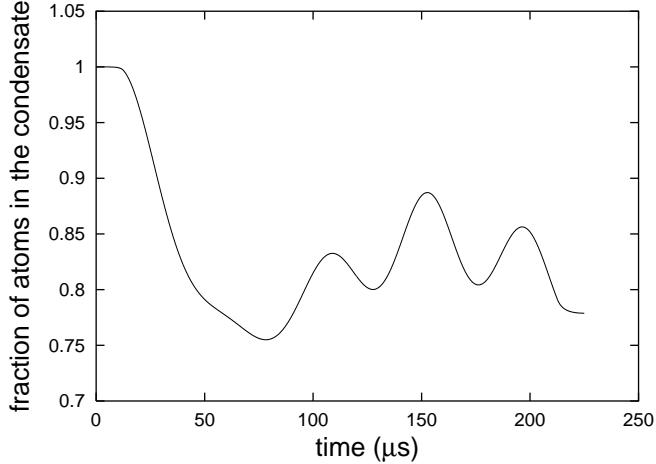}
\caption{ \label{nc_treal}
     The fraction of atoms in the condensate as a function of time for a
     calculation that includes only quantum evaporation. The initial
     number of condensate atoms is $N_{\text{c}} (0)=16500$. The rise time
     $t_{\text{rise}}=12.5$ $\mu$s and the hold time is
     $t_{\text{hold}}=200$ $\mu$s. The scattering length is equal to $a=2000
     a_0$ during hold.
  }
\end{figure}

\subsubsection{Results}
We perform our calculations for the parameters of the experiment by
Claussen \textit{et al.} \cite{neil}. In particular, the frequencies of
the external trapping potential are the same as in the previous section.
Fig.~\ref{nc_treal} shows the fraction of atoms in the condensate as a
function of time, for a pulse such that $t_{\text{rise}}=12.5$ $\mu$s and
$t_{\text{hold}}=200$ $\mu$s. The magnetic field during the hold is
$B=156.9$ G, which corresponds to a scattering length of $a=2000 a_0$. The
simulation shows that once the scattering length nearly takes on its
largest value and the coupling between the condensate and the excited
states is therefore
largest, the condensate starts ejecting atoms. Part of these atoms then
oscillate back and forth between the condensate and the excited states.
The curve in Fig.~\ref{nc_treal} clearly contains several frequencies
since we are dealing with several excited states and thus a multimode Rabi
oscillation. At the end of the pulse the rate, i.e., the slope of the
curve, becomes equal to zero because the coupling between the condensate
and the excited states becomes equal to zero at the end of the
pulse, where the scattering length is equal to zero.

\begin{figure}
\includegraphics{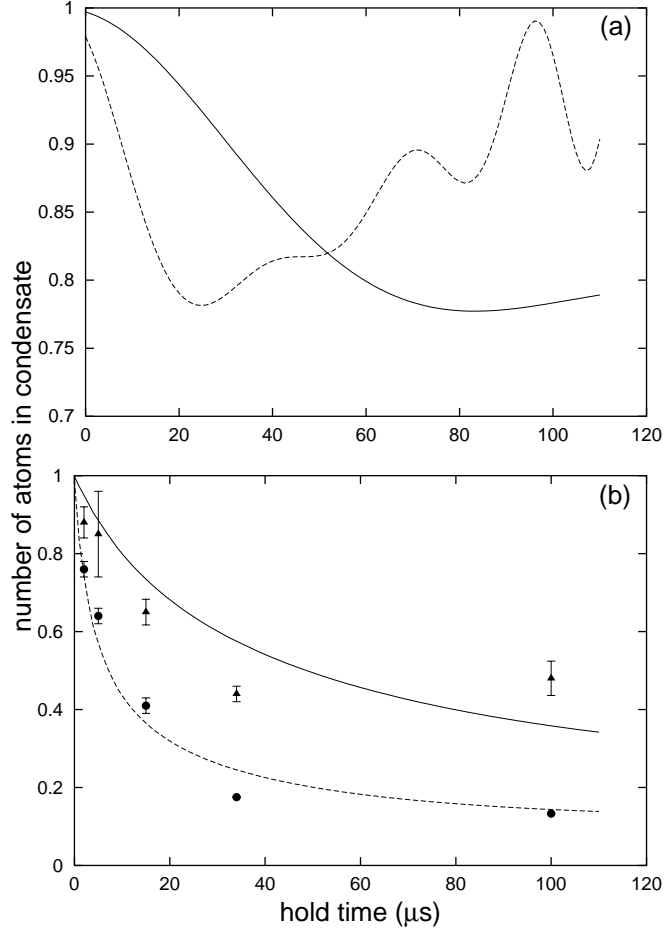}
\caption{ \label{nc_thold2}
    Fraction of atoms in the condensate as a function of the hold time.
  The rise time is kept fixed at the value $t_{\text{rise}}=12.5$ $\mu$s.
  The initial number of condensate atoms is taken  $N_{\text{c}} (0)=6100$
  (solid line) and $N_{\text{c}} (0)=16500$ (dotted line). The scattering
  length is equal to $a=2000 a_0$ during hold. (a) The result of the
  calculation that includes only quantum evaporation. (b) The calculation
  that includes both quantum evaporation and three-body recombination. The
  experimental points are taken from Ref.~\protect\cite{neil}. }
\end{figure}
To compare our results with the available experimental data we calculate
the number of atoms as a function of the hold time $t_{\text{hold}}$ and
the rise time $t_{\text{rise}}$. Fig.~\ref{nc_thold2}~(a) shows the result
of a calculation of the fraction of condensate atoms as a function of
$t_{\text{hold}}$, with $t_{\text{rise}}=12.5$ $\mu$s fixed. The
calculation is done for two different initial numbers of condensate atoms.
The solid line displays the result for $N_{\text{c}} (0)=6100$ and the dashed
line for $N_{\text{c}} (0)=16500$. Notice that the latter initially has a larger
slope because the effective Rabi coupling between the condensate
and the excited states is larger in this case. This is because of the fact
that it is proportional to the condensate density.

The results of the simulation that includes only quantum evaporation,
shown in Fig.~\ref{nc_thold2}~(a), show an oscillation in the fraction of
condensate atoms as a function of the hold time. This oscillation is not
observed in experiment, because of the fact that three-body recombination
plays an important role in this case since it becomes large near the
resonance \cite{jake2000}. Therefore, we also want to perform a
calculation that includes both quantum evaporation and three-body
recombination. However, the magnetic-field dependence of the rate constant
for this process is unknown. Nevertheless, we are able to make progress by
realizing that the hold time is generally larger than the rise time for
the experimental points shown in Fig.~\ref{nc_thold2}. Since
experimentally three-body recombination is known to increase by orders of
magnitude near the resonance \cite{jake2000}, the contribution of
three-body recombination will be most important during hold, where the
magnetic field is closest to the resonance. This suggests that we only
need to include it during hold. Note that for this approximation to be
valid it is essential that the rise time is shorter that the hold time. If
the rise time is larger than the hold time the magnetic field dependence
of the three-body recombination rate constant is of importance, since the
magnetic field is then time dependent for almost the entire pulse.

Fig.~\ref{nc_thold2}~(b) shows the result for a calculation that includes
both quantum evaporation and three-body recombination, with a rate
constant $K_3=3 \times 10^{-23}$ cm$^6$/s during hold. This value for the
normal-component rate constant agrees with the order of magnitude of the
experimental data \cite{jake2000}. This calculation shows good
quantitative agreement with experiment for both initial numbers of
condensate atoms.

\begin{figure}
\includegraphics{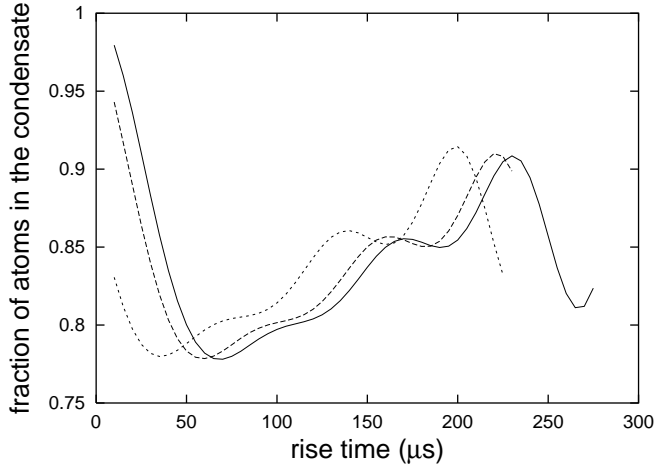}
\caption{ \label{nc_trise2}
     Fraction of atoms in the condensate as a function of the rise time
    for a calculation that includes only quantum evaporation. The hold
    time is kept fixed at $t_{\text{hold}}=1$ $\mu$s (solid line),
    $t_{\text{hold}}=5$  $\mu$s (dashed line), and $t_{\text{hold}}=15$
    $\mu$s (dotted line). The scattering length is equal to $a=2000 a_0$
    during hold.
  }
\end{figure}

Finally, we have calculated the number of atoms in the condensate
as a function of the rise time. The result of this calculation is
shown in Fig.~\ref{nc_trise2}, for various hold times. The solid
line corresponds to $t_{\rm hold}=1$ $ \mu$s. The dashed and
dotted line correspond to hold times of $5$ $ \mu$s and $15$ $
\mu$s, respectively. The scattering length during hold is equal to
$a=2000 a_0$ for this simulation. For all the results in
Fig.~\ref{nc_trise2} the rise time of the pulse is larger than
the hold time. This means that the magnetic-field dependence of
the three-body recombination rate constant is very important in
this case, since the magnetic field is varying most of the time.
Fitting the dependence to the experiments is difficult due to the
long times taken by the numerical computations. Therefore we
refrain from including three-body recombination in these
simulations. Nevertheless, there is agreement with the
experimental results regarding several aspects of our results.
First, we find that the number of atoms increases with the rise
time over some interval. This was also found in the experiment by
Claussen \textit{et al.} \cite{neil}. Note that this fact can not
be explained by any loss process characterized by a rate constant
because the amount of atom loss will then always be larger with
longer times. Second, the minima of the curves in
Fig.~\ref{nc_trise2} shift to lower values of $t_{\rm rise}$ with
longer hold times. This was also observed in the experiment by
Claussen \textit{et al.} \cite{neil}. These minima also occur on
approximately the experimental values of $t_{\rm rise}$. The fact
that in the experiment the minima become lower with increasing
hold time can be explained by three-body recombination.

In conclusion, we have applied the generalized Gross-Pitaevskii
equation  to the recent single-pulse experiments by Claussen
\textit{et al.} \cite{neil}. We have shown that the number of
atoms increases with time over some ranges. This can not be
understood in terms of conventional loss processes such as
three-body recombination or dipolar decay. However, to obtain
agreement with the available experimental data we had to include
three-body recombination in our simulations. Due to the fact that
the magnetic field dependence of this process is completely
unknown we are not able to make a fit to experiment in all
situations.

\subsection{Atom-molecule coherence}
\begin{figure}
\includegraphics{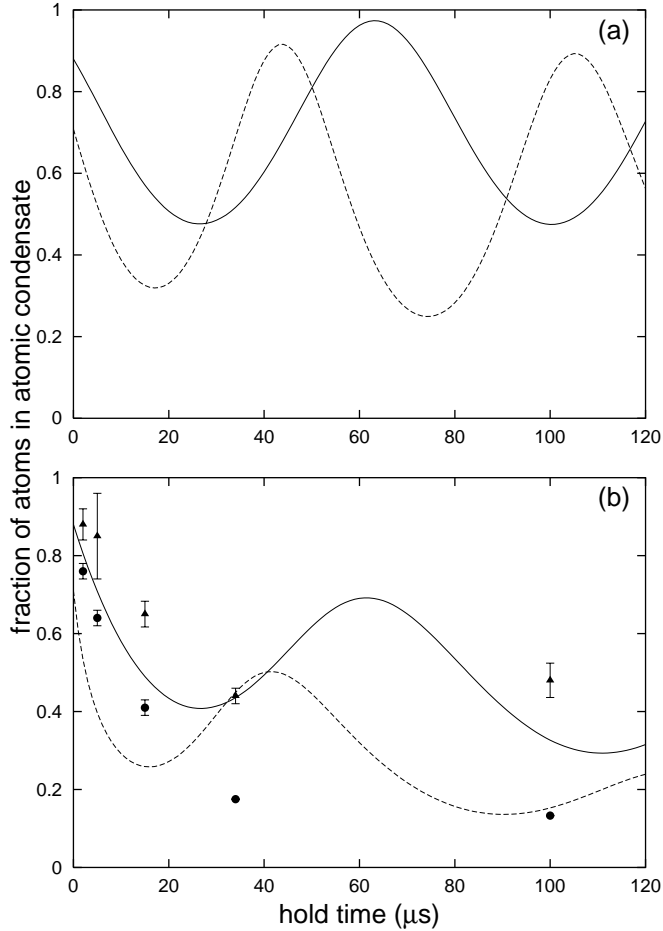}
\caption{\label{ncth_molecules}
  Fraction of atoms converted in the condensate as a function of the hold time.
  The rise time is kept fixed at the value $t_{\text{rise}}=12.5$ $\mu$s.
  The initial number of condensate atoms is taken  $N_{\text{c}} (0)=6100$
  (solid line) and $N_{\text{c}} (0)=16500$ (dashed line). The scattering
  length is equal to $a=2000 a_0$ during hold. (a) The result of the
  calculation that includes only the coupling of the atomic condensate to
  the molecular field. (b) The calculation
  that includes both atom-molecule coupling and three-body recombination. The
  experimental points are taken from Ref.~\protect\cite{neil}.
   }
\end{figure}
Recent experimental and theoretical work has shown that atom-molecule
coherence is of importance in the case of a double pulse in the magnetic
field \cite{elisabeth2,servaas,kohler2002} . Therefore, we may expect it to
have an important effect in the case of the single-pulse experiments as well. To make
the discussion of these experiments more complete, we investigate the role
of the molecules by means of a quantum field theory that we derived recently
\cite{rembert4}. This theory incorporates the correct
molecular binding energy and scattering properties of the atoms at the
quantum level by using coupling constants that are dressed by ladder
diagrams and by including the molecular self-energy.
Introducing Heisenberg operators $\hat \psi_{\rm a}$ and $\hat \psi_{\rm
m}$ that annihilate an atom and a bare molecule, respectively, the hamiltonian
for the atom-molecule system reads
\begin{eqnarray}
\label{eq:heom}
  i \hbar \frac{\partial \hat \psi_{\rm m} \args}{\partial t}
    &=&\left[  -\frac{\hbar^2 {\bf \nabla}^2}{4m}
            +\delta (B(t))  - g^2 \frac{m^{3/2}}{2 \pi \hbar^3} i
  \sqrt{i \hbar \frac{\partial}{\partial t}
    +\frac{\hbar^2 {\bf \nabla}^2}{4m}}~
    \right] \hat \psi_{\rm m} \args + g \hat \psi_{\rm a}^2
    \args~,\nonumber \\
     i \hbar \frac{\partial \hat \psi_{\rm a} \args}{\partial t}
    &=&\left[  -\frac{\hbar^2 {\bf \nabla}^2}{2m}
            + T^{\rm 2B}_{\rm bg} \hat \psi_{\rm a}^{\dagger}
        \args \hat \psi_{\rm a} \args
    \right] \hat \psi_{\rm a} \args 
    + 2 g \hat \psi_{\rm a}^{\dagger}
          \args \hat \psi_{\rm m} \args~. 
\end{eqnarray}
Here, $g=\hbar \sqrt{2 \pi a_{\rm bg} \Delta B  \Delta \mu /m}$ is the
atom-molecule coupling constant and \mbox{$\delta (B) = \Delta \mu
(B(t)-B_0)$}
denotes the detuning, i.e., the energy difference between two
atoms and the bare molecule. It is determined by the difference in magnetic
moment between the atoms and the bare molecule, which in the case of $^{85}$Rb
is equal to $\Delta \mu \approx -2.2 \mu_{\rm B}$ \cite{servaas}.

At first glance the term proportional to $\sqrt{i \hbar \partial/\partial t +
\hbar^2 \nabla^2 /(4m)}$ may appear unexpected. It corresponds to the
imaginary part of the self energy of the bare molecule which arises
physically from the fact that the molecular state interacts with the
two-atom continuum. This affects both the wave function of the dressed molecule
and its binding energy. By determining the pole of the molecular
propagator for negative detuning, the latter can be shown to be given by
\cite{rembert4}
\begin{equation}
\label{eq:ebres}
  \epsilon_{\rm m} (B)= \delta (B) + \frac{g^4 m^3}{8 \pi^2 \hbar^6}
   \left[\sqrt{1-\frac{16 \pi^2 \hbar^6}{g^4 m^3} \delta (B)} -1\right]~,
\end{equation}
which reduces to $\epsilon_{\rm m} (B) = -\hbar^2/[m (a(B))^2]$ for values
of the magnetic field close to the resonance. Due to the coupling with the
continuum of atoms, i.e., the open channel of the Feshbach problem, the
molecular state is strongly affected and is given by
\begin{equation}
\label{eq:wavefctmol}
  | \chi_{\rm m}; {\rm dressed} \rangle=  
                 \sqrt{Z(B)}|  \chi_{\rm m} ; {\rm bare} \rangle
                + \int \frac{d {\bf k}}{(2 \pi)^3} C({\bf k})
                           | {\bf k}, -{\bf k};{\rm open} \rangle ~,
\end{equation}
where the coefficients $C({\bf k})$ are normalized as $\int d
\bk |C(\bk)|^2/(2 \pi)^3  = 1-Z(B)$. It
contains with an amplitude $\sqrt{Z(B)}$ the bare molecular state
$|\chi_{\rm m};{\rm bare} \rangle$. Moreover, because of the coupling to the
two-atom continuum the molecule acquires a nonzero component in the open
channel. The wave function renormalization factor $Z(B)$ is given by
\cite{rembert4}
\begin{equation}
 Z(B)=\frac{1}{1 + g^2 m^{3/2}/(4 \pi \hbar^3 \sqrt{|\epsilon_{\rm m}(B)|})}~,
\end{equation}
which approaches one for values of the magnetic field far off resonance,
where the dressed molecular state reduces to the bare molecular state, as
expected. However, for values of the magnetic field close to the
resonance, it is much smaller than one. In particular, for the case of the
single-pulse experiments we always have that $Z(B) \ll 1$, which implies
that the magnetic moment of the dressed molecule is in very good approximation
equal to twice the magnetic moment of an atom. For magnetically trapped
atoms this implies that the dressed molecule is subject to twice the trapping
potential for the atoms. With respect to this remark it is important to
note that the result of the calculations of Kokkelmans and Holland
\cite{servaas} for the density of the molecular condensate should be
multiplied by a factor $1/Z(B) \gg 1$ to obtain the density of real
dressed molecules, since these authors calculate the expectation value
of the bare molecular field operator $\langle \hat \psi_{\rm m} \args
\rangle$.  

\begin{figure}
\includegraphics{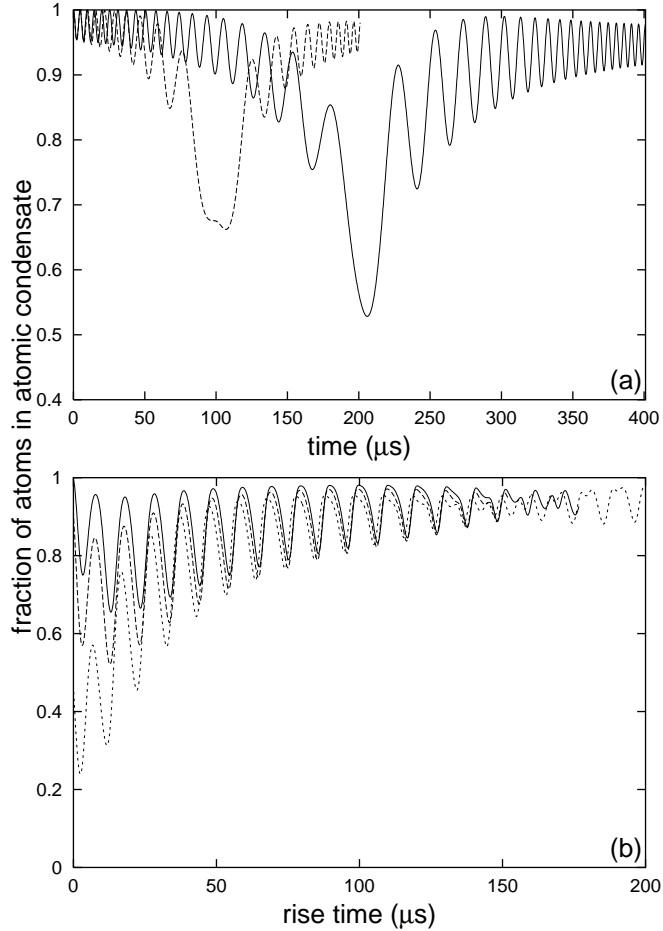}
\caption{\label{nctr_molecules}
   Fraction of atoms in the atomic condensate for the situation where the
   atomic condensate is coupled to the molecular field. Initially there
   are $N_{\rm c} (0)=16500$ atoms and no molecules. The magnetic field is
   such that $a=2000 a_0$ during the hold. (a) Fraction of atoms as a
   function of the real time for $t_{\rm rise}=200$ $\mu$s (solid line)
   and $t_{\rm rise}=100$ $\mu$s (dashed line). The hold time is equal to
   $t_{\rm hold}=1$~$\mu$s for both pulses. (b) Fraction of atoms as a
   function of the rise time for different hold times of $t_{\rm hold}=1$
   $\mu$s (solid line), $t_{\rm hold}=5$ $\mu$s (dashed line) and $t_{\rm
   hold}=15$~$\mu$s (dotted line).
   }
\end{figure}
To bring out the physics of Eq.~(\ref{eq:heom}) more clearly, we introduce
the operator $\hat \psi'_{\rm m} =\hat \psi_{\rm m}/\sqrt{Z(B)} $ that
creates a dressed molecule, i.e., a molecule with an internal state as in
Eq.~(\ref{eq:wavefctmol}). Since we intend to consider the situation where
initially all atoms are in the atomic condensate, we are allowed to make a
mean-field approximation for the atomic field operator and consider only its
expectation value. There are however no molecules present at the initial
time and this requires a quantum treatment of the molecular field
operators. The resulting equations for the atomic condensate wave function
coupled to the dressed molecular field reads for the experimental
conditions of interest
\begin{eqnarray}
\label{eq:mfeqs}
   i \hbar \frac{\partial \phi_{\rm a} \args}{\partial t}
    &=&\left[  -\frac{\hbar^2 {\bf \nabla}^2}{2m}
            + V^{\text{ext}} (\bx)+ T^{\rm 2B}_{\rm bg} |\phi_{\rm a}
        \args|^2 
    \right] \phi_{\rm a} \args 
    + 2 g \sqrt{Z(t)} \phi^*_{\rm a} \args \hat \psi'_{\rm m} \args~, \nonumber \\
  i \hbar \frac{\partial \hat \psi'_{\rm m} \args}{\partial t}
    &=&\left[  -\frac{\hbar^2 {\bf \nabla}^2}{4m}
            + 2 V^{\text{ext}} (\bx) +\epsilon_{\rm m} (t) 
    \right] \hat \psi'_{\rm m} \args + g \sqrt{Z(t)} \hat \psi_{\rm a}^2
    \args~,\nonumber \\
\end{eqnarray}
where $\phi_{\rm a} \equiv \langle \hat \psi_{\rm a} \rangle$, $Z(t) \equiv
Z(B(t))$, and $\epsilon_{\rm m} (t) \equiv \epsilon_{\rm m} (B(t))$. 
In the derivation of the above coupled equations we have assumed that we
are allowed to make an adiabatic approximation for the renormalization
factor $Z(B)$ and that we can evaluate it at every time at the magnetic field
$B(t)$. In principle there are retardation effects due to the fact that
the dressed molecular state does not change instantaneously. It turns out
that these effects can be neglected if 
\begin{equation}
\label{eq:condition}
  \hbar \left| \frac{\partial \ln Z(B(t))}{\partial t} \right| \ll | \epsilon_{\rm m}
  (B(t))|~,
\end{equation}
which is fulfilled for almost the entire duration of most of the
pulses in the experiments of Claussen \textit{et al.} \cite{neil}. 
We come back to this point in the discussion at the end of the paper. In
principle, the coupling between the two-atom continuum and the molecule
also contains an incoherent part corresponding to the rogue dissociation
process considered by Mackie {\it et al.} \cite{mackie2002}. The rate for
this process will be small, however,  under the condition given in
Eq.~(\ref{eq:condition}). Moreover, the mean-field effects of the
condensate on the thermal atoms will suppress this process even further.
It can, in principle, be included straightforwardly and will take the form of a
dissipation term in the equation for the molecular operator.

We solve the equations for the atomic condensate wave function coupled to
the dressed molecular field by using for the condensate wave function again the
gaussian \textit{ansatz} in Eq.~(\ref{gaussianansatz}), and by expanding
the dressed molecular annihilation operator in harmonic oscillator
eigenstates, similar to the expansion in Eq.~(\ref{expansionoperator}). As
initial condition we assume that at $t=0$ only condensed atoms are
present. The results of our calculations are shown in
Figs.~\ref{ncth_molecules}~and~\ref{nctr_molecules}. 

The calculations presented in Fig~\ref{ncth_molecules} are performed for
the same experimental conditions as in Fig.~\ref{nc_thold2}. This result
clearly shows that a large fraction of atoms is coherently converted into
molecules as a result of the fast ramp in the magnetic field and that
these oscillate back and forth between the atomic condensate and the
molecular states. Due to the fact that the conversion of the atoms to the
molecular states is coherent, the operator $\hat \psi'_{\rm m} \args$
acquires a nonzero expectation value. Fig~\ref{ncth_molecules}~(b) 
shows the results of simulations where also three-body decay is taken into
account in the same manner as in the previous section. The normal
component rate constant is taken equal to $K_3 = 3 \times 10^{-23}$
cm$^6$/s. Interestingly, the initial decay without three-body
recombination is already larger than the experimental data and by adding
three-body recombination it is therefore impossible to make a fit to the
experimental data. This is possibly the result of neglecting the
retardation effects of the renormalization factor $Z(B(t))$ and the
rogue-dissociation process, since the condition in
Eq.~(\ref{eq:condition}) is violated for a significant fraction of the
total duration of the pulse in this case. For the simulations presented in
Fig.~\ref{nctr_molecules} this condition is violated only for a very small
fraction of the total duration of the pulse for rise times larger that
$t_{\rm rise} \approx 50$~$\mu$s and is not violated at all for $t_{\rm rise}
\geq 150$~$\mu$s. Note that the effect of retardation and rogue
dissociation lead to decoherence, which means that our calculations give an
upper bound on the amount of molecules that are coherent with the atoms.

Fig.~\ref{nctr_molecules}~(a) shows the result of two calculations for
different rise time as a function of the total time. As expected, the
number of atoms in the atomic condensate first oscillates with large
frequency since the dressed molecular binding energy is large here. As the
magnetic field approaches values closer to the resonance, the frequency
decreases. From Fig.~\ref{nctr_molecules}~(a) it is clear that only the
largest frequency, which also has the largest amplitude since the gas is
then closest to resonance, gives a significant contribution to the
frequency observed in the number of condensate atoms as a function of the
rise time, because the larger frequencies with smaller amplitude
average out. However, these oscillations are not observed in the
experimental data of Ref.~\cite{neil}. Introducing three-body
recombination to fit the theory to experiment is impossible with pulses having
relatively long rise times, for the same reasons as in the previous
section. Nevertheless, the amplitude of the oscillations are in this case,
except for the longest hold time,  comparable to that of the
simulations where only quantum evaporation is included. This implies that
for a thorough treatment of the single-pulse experiments both
atom-molecule coherence and quantum evaporation should be included. This
is beyond the scope of the present paper but work in this direction is in
progress.  

\section{Conclusions}
\label{conclusions} We have put forward a generalized Gross-Pitaesvkii
equation that includes nonlocal terms which describe the quantum
evaporation of the Bose-Einstein condensate. We have applied this equation
to two experimental situations which make use of a Feshbach resonance to
alter the interaction properties of the atoms. First we have considered
the case where the condensate undergoes a Type~II~collapse whose dynamics
is mainly determined by the external trapping potential and have found
good quantitative agreement with experiment. Second we have considered the
recent single-pulse experiments \cite{neil}. In general we have also found
agreement with experiment in this case, keeping in mind that the
magnetic-field dependence of the three-body recombination rate constant is
completely unknown. The latter is the first serious complication in the
theoretical analysis. Apart from considering quantum evaporation we have
also considered the role of atom-molecule coherence in the single-pulse
experiments, by means of an adiabatic approximation to our effective
quantum field theory for the description of Feshbach resonances
\cite{rembert4}. In first instance atom-molecule coherence appears to be
an important effect. However, the second theoretical complication is that
the adiabatic approximation in general overestimates the effect and does
not take into account rogue dissociation \cite{mackie2002}. Including this
process damps out the Rabi oscillations between the atoms and molecules
and leads to the production of energetic atoms that may contribute to the
experimentally observed bursts \cite{elisabeth2}. Due to these two
complications, a completely satisfying quantitative description of these
experiments is still lacking.  It should be mentioned that our
calculations take into account the inhomogeneity of the trapped gas
exactly and not in local-density approximation. In addition, we do not
make a single-mode approximation either for the atomic condensate or the
dressed molecules. In future work we intend to consider quantum
evaporation, rogue dissociation, and three-body recombination
simultaneously to obtain more insight into these intriguing JILA
experiments.

\begin{acknowledgments}
It is a pleasure to thank Neil Claussen, Eric Cornell, Simon Cornish,
Elisabeth Donley, and Carl Wieman for helpful remarks that have contributed to
this paper.

\end{acknowledgments}

\appendix
\section{Rate equation for Type~II~collapse}
\label{app:rateeqn}
With the gaussian \textit{ansatz} for the condensate wave function and the
semiclassical propagator for the ejected atoms, the final rate equation for the
change in the number of atoms is given by
\begin{eqnarray}
  \frac{dN_{\text{c}}(t)}{dt} &=& -64 i \sqrt{6} \hbar^3 a(t) N^{3/2}_{\text{c}}
  (t)
   \int_{-\infty}^t dt' \left\{ \rule{0mm}{9mm} a(t') N_{\text{c}}^{3/2} (t')
        e^{-i (\theta_0 (t)-\theta_0 (t'))-\frac{2i}{\hbar} |\phi ({\bf 0},
    (t+t')/2)|^2 (t'-t)} \right. \nonumber \\
     &&
       \times \left[ \rule{0mm}{7mm}
         \left\{
       m \pi q_r (t) q_r(t')
        \left[ i q_r (t) q_r (t') \left( (t-t') \dot q_r (t) - q_r (t)
	\right)
                                \dot q_r(t')
          -q_r (t') \dot q_r (t)
        \right] m^2 \right. \right.\nonumber \\
     && \left.
    + 3 m \hbar \left[ q^2_r (t) +
                  (t'-t) (\dot q_r (t) q_r (t) - q_r (t') \dot q_r (t'))
            +q_r^2 (t')
              \right] + 9 i \hbar^2 (t'-t)
     \right\} \nonumber \\
     && \times
      \sqrt{\frac{i q_z (t) \dot q_z (t')}{\omega_z} +3}~
      \sqrt{q_z (t) q_z (t') \left( 3 - \frac{i m q_z (t') \dot q_z
      (t')}{\hbar }\right)} \nonumber \\
     && \times \left. \left.
         \sqrt{
           \frac{m \dot q_z (t) q^3_z (t)}{3 i \hbar m q_z (t) \dot
           q_z (t) }
           + q^2_z (t)
           + 3 i \hbar
           \left(
             \frac{q_z^2 (t')}{3 i \hbar + m q_z (t') \dot q_z (t')}
         +\frac{t-t'}{m}
           \right)
          }
       \right]^{-1}
     \right\} \nonumber \\
     &&- K_3 \frac{N^{3}_{\text{c}}(t)}{3 \sqrt{3} \pi^3 q^4_r (t) q^2_z
     (t)}.
\end{eqnarray}
Here, $q_r (t)$ and $q_z (t)$ denote the radial and axial width of the
condensate, respectively.

\section{Matrix elements}
\label{app:matrixelnts} In this Appendix we calculate the matrix
elements $V_{\bf n,n'} (t)$ in Eq.~(\ref{matelnts}). Because of
the fact that we are dealing with a trapping potential  that is
symmetric around the $z$-axis, the excited states factorize into
a radial and an axial part. It is convenient to characterize the
radial part of the excited states by the quantum numbers
$(n_r,m)$, where $n_r$ counts the number of radial nodes in the
wave function and $m$ is the quantum number corresponding to the
projection of the angular momentum on the $z$-axis. The third
quantum number $n_z$ counts the number of nodes in the axial
direction. In cylindrical coordinates these states are given by
\cite{bj}
\begin{eqnarray}
\label{statesfull}
  \chi_{n_r,m,n_z} (r,\theta,z) &\propto& e^{-r^2/(2 l_r^2)} |r|^m
      \mbox{$_1$}F_1(-n_r,|m|+1,(r/l_r)^2) e^{i m \theta} \nonumber \\
     && \times H_{n_z} (z/l_z) e^{-z^2/(2 l_z^2)},
\end{eqnarray}
where $l_i \equiv \sqrt{\hbar/(m \omega_i)}$. The Hermite polynomials are
denoted by $H_n (x)$ and the confluent hypergeometric function is
denoted by $\mbox{$_1$}F_1 (p,q,x)$. The overlap integral with two
functions of the form as in Eq.~(\ref{statesfull}) with a
gaussian of arbitrary width is, to the best of our knowledge, not
tabulated. Nevertheless, we can make analytical progress by
realizing that we only have to take into account the states with
$m=0$, since the interaction conserves parity.  The radial part
of these states is given by
\begin{eqnarray}
\label{radialpart}
  | \chi_{n_r} \rangle = \frac{1}{n_r !}
          \left( \frac{\hat a_x^{\dagger} - i \hat a_y ^{\dagger}}{\sqrt{2}} \right)^{n_r}
          \left( \frac{\hat a_x^{\dagger} + i \hat a_y ^{\dagger}}{\sqrt{2}} \right)^{n_r}
      | 0 \rangle,
\end{eqnarray}
where the operator $(\hat a_x^{\dagger} - i \hat a_y ^{\dagger})/\sqrt{2}$
lowers the magnetic quantum number $m$ of the angular
momentum by one. The operator $(\hat a_x^{\dagger}
+ i \hat a_y ^{\dagger})/\sqrt{2}$ raises this quantum number by one. Here,
the
operators $\hat a_i^{\dagger} \equiv \sqrt{m \omega_i/(2 \hbar)}(\hat x_i
- i \hat p/(m \omega_i))$ are the usual harmonic-oscillator creation
operators. The ground state is denoted by $| 0 \rangle$. The creation operators
commute and hence we can rewrite the radial wave function of the state as
\begin{eqnarray}
  | \chi_{n_r} \rangle = \sum_{n=0}^{n_r} \frac{\sqrt{(2n)!}
  \sqrt{(2(n_r-n))!}}{n!(n_r-n)!2^{n_r}} | 2n \rangle_x |2m \rangle_y,
\end{eqnarray}
where $| n \rangle_i$ denote the normalized eigenstates of the
hamiltonian $H_i = \hat p_i^2/(2m)+m \omega_i \hat x_i^2/2$ of
the one-dimensional harmonic oscillator. In the derivation of
this expression we used Newton's binomium to rewrite the $n_r$-th
powers of the operators in the right-hand side of
Eq.~(\ref{radialpart}) as a sum of $n_r$ terms. With this result,
the normalized wave functions of the excited states of interest
are given by
\begin{eqnarray}
  \chi_{n_r,n_z} (\bx) &=&
  \frac{1}{4^{n_r} n_z! 2^{n_z} \pi} \left[ \sum_{n=0}^{n_r}
  \frac{1}{n! (n_r-n)!}
  e^{-(x^2+y^2)/(2 l_r^2)} H_{2n} (x/l_r) H_{2(n_r-n)} (y/l_r) \right]
  \nonumber \\ && \times e^{-z^2/(2 l_z^2)} H_{n_z} (z/l_z),
\end{eqnarray}
The overlap integrals of two excited states of this form with a
gaussian of arbitrary width are tabulated \cite{tables}. The
final result for the matrix elements is given by
\begin{eqnarray}
  V_{n_r,n_z;m_r,m_z} (t)&=&  \frac{N_{\rm c} (t)}{\pi^{3/2}
  l_r^2 q_z (t) \left((q_r(t)/l_r)^2 +1\right)}  \sqrt{\frac{q^2_z(t)}{q_z^2(t)+l_z^2}}
 \nonumber \\
  && \times  \frac{1}{2^{n_r+m_r}}  \left[ \sum_{n=0}^{n_r} \sum_{m=0}^{m_r}
                      \sum_{k=0}^{\min (2n,2m)}
              \ \sum_{l=0}^{\min(2(n_r-n),2(m_r-m))} \right. \nonumber \\
  &&
  \frac{1}{n! (n_r-n)!} \frac{1}{m! (m_r-m)!}
  \frac{(2n)!}{(2n-k)!} \frac{2 (n_r-n)}{(2 (n_r-n)-l)!}
  \nonumber \\
  && \times
     \left(\begin{array}{c}
     2m \\
     k
    \end{array} \right)
    \left( \begin{array}{c}
     2(m_r-m) \\
     l
    \end{array} \right) (2(m+n-k)-1)!! \nonumber \\
    && \times \left. (2(n_r+m_r-n-m-l)-1)!!
    \ ((q_r(t)/l_r)^2+1)^{k+l-n_r-m_r}
  \rule{0mm}{7mm} \right] \nonumber \\
  && \times \left[ \sum_{q}^{\min(n_z,m_z)}  \frac{\sqrt{n_z!m_z!}}{q!(n_z-q)!(m_z-q)!}
       \right. \nonumber \\
    && \left. \times
       ((q_z(t)/l_z)^2+1)^{q/2-(m_z+n_z)/4} (n_z+m_z-2q-1)!! \rule{0mm}{7mm} \right],
\end{eqnarray}
if $n_z+m_z$ is even, otherwise it is equal to zero.

\end{document}